\documentclass[preprints,article,accept,moreauthors,LaTeX,dvi2pdf]{mdpi}

\firstpage{1}
\pubvolume{9}
\issuenum{1}
\articlenumber{10}
\pubyear{2021}
\copyrightyear{2021}
\externaleditor{Academic Editor: Alberto C. Sadun} 
\datereceived{6 January 2021}
\dateaccepted{23 January 2021}
\datepublished{26 January 2021}
\hreflink{\textls[-25]{http://doi.org/10.3390/galaxies9010010}} 
\usepackage[export]{adjustbox}

\Title{Powerful Jets from Radiatively Efficient Disks, a Decades-Old Unresolved Problem in High Energy Astrophysics}

\TitleCitation{\textls[-25]{Powerful Jets from Radiatively Efficient Disks, a Decades-Old Unresolved Problem in High Energy Astrophysics}}

\Author{{Chandra B. Singh} 
$^{1,}$*$^{,\dagger}$, {David Garofalo} 
$^{2,\dagger}$\href{https://orcid.org/0000-0001-5536-829X}{\orcidicon} and {Benjamin Lang} 
$^{3}$}

\AuthorNames{Chandra B. Singh, David Garofalo and Benjamin Lang}

\AuthorCitation{Singh, C.B.; Garofalo, D.; Lang, B.}

\address{
$^{1}$ \quad South-Western Institute for Astronomy Research, Yunnan University, University Town, Chenggong, Kunming~650500, China\\
$^{2}$ \quad Department of Physics, Kennesaw State University, Marietta, GA 30060, USA; dgarofal@kennesaw.edu\\
$^{3}$ \quad Department of Physics \& Astronomy, California State University, Northridge, CA 91330, USA; benjamin.lang.447@my.csun.edu}

\corres{Correspondence: chandrasingh@ynu.edu.cn}

\firstnote{These authors contributed equally to this work.}

\abstract{The discovery of 3C 273 in 1963, and the emergence of the Kerr solution shortly thereafter, precipitated the current era in astrophysics focused on
using black holes to explain active galactic nuclei (AGN). But while partial success was achieved in separately explaining the bright nuclei of some AGN via thin disks, as well as powerful jets with thick disks, the combination of both powerful jets in an AGN with a bright nucleus, such as in 3C 273, remained elusive. Although numerical simulations have taken center stage in the last 25 years, they have struggled to produce the conditions that explain them. This is because radiatively efficient disks have proved a challenge to simulate. Radio quasars have thus been the least understood objects in high energy astrophysics. But recent simulations have begun to change this. We explore this milestone in light of scale-invariance and show that transitory jets, possibly related to the jets seen in these recent simulations, as some have proposed, cannot explain radio quasars. We then provide a road map for a resolution.}


\begin{document}

\section{Introduction}
Radio quasars are active galaxies that harbor strong, collimated jets, mostly of FRII 
radio morphology ~\cite{1}
, 
but that also shines brightly in the optical, suggesting they are accreting black holes in radiatively efficient mode~\cite{2}. They are not distributed randomly in space and time, with a greater probability of being found in isolated environments and at higher redshifts, at least compared to FRI 
radio galaxies, which are found in richer environments and comparably lower redshifts~\cite{3,4}. FRI radio galaxies differ from radio quasars not only in their jet morphology but also in their ‘radio mode' accretion, as opposed to `quasar mode', indicating the disk is radiatively inefficient.

Many X-ray binaries cycle through states with radio-emitting jets present
during so-called `hard' states, but absent during so-called ‘soft’ states~\cite{5,6}. The consensus is that the former is related to radiatively inefficient accretion, similar to jet mode accretion in FRI radio galaxies, while the latter is associated with radiatively efficient
accretion, similar to quasar mode (\cite{7,8,9,10} and references therein). It is often found that as X-ray binaries transition from the hard state to the soft one, a transitory, ballistic, or more powerful and collimated jet, is generated~\cite{11,12}. It has been claimed that such a jet taps into black hole rotational energy~\cite{13} unlike the hard state jet~\cite{9}. Despite these
general properties, a subclass of X-ray binaries does not suppress jets in soft
states~\cite{14}. And when these neutron star X-ray binaries do
suppress jets, their transitory jets tend to be weaker~\cite{15}.

The observational evidence described above points to a picture in which accreting supermassive black holes combine radiatively efficient accretion with jets while stellar-mass black holes do not. In the past, and again recently via general relativistic magnetohydrodynamic (GRMHD) simulations, resolving this issue hinges on the assumption that radio quasars are the large scale analog of transitory burst states in black hole X-ray binaries, during which X-ray binaries are approaching the soft state, so a stable radiatively efficient disk is in the process of formation, although it has not yet formed. The goal of this work is to highlight the incompatibility between well-accepted theory and the observations.
We then produce a roadmap for GRMHD simulations designed for getting at a better solution,
one that appeals to ideas that have been available for a decade.

In Section \ref{sec2}, we analyze the history of GRMHD simulations in their struggle to generate thin disks in an attempt to understand the disk radiative efficiency. The point here is to understand the conditions that produce jet suppression. Within the context of that history, we single out recent simulations by~\cite{16,17} that promise to take our understanding of jet formation in thin disks to a new level. This leads to an exploration of the jet-disk connection in light of scale invariance and the roadmap alluded above. We then conclude.

\section{Disks and Jets in GRMHD}
\label{sec2}
Within the last half-decade, GRMHD simulations have pushed the boundary of disk thickness to extremes, with values about an order of magnitude smaller than two decades ago. The height to radius ratio, or scaleheight, ${H/R}$, is found to be on average equal to 0.20--0.26 in the simulations dating back a decade and a half ago~\cite{18,19,20,21,22,23,24}, while, a decade later, we see $H/R$ decrease to 0.13~\cite{25}. In-between we see a disk scaleheight of $H/R \sim 0.1$ associated with disk radiative efficiency of $6\%$ that of the relativistic Novikov-Thorne model~\cite{26}, followed up with disks as thin as $H/R \sim 0.06$ but up to $10\%$ difference in efficiency with respect to the Novikov \& Thorne disk~\cite{27,28}. Similar results were found in~\cite{29}, with $H/R \sim 0.1$ and deviation from Novikov \& Thorne disks of $5\%$. By 2019 we see $H/R \sim 0.03$~\cite{16,17}. To model radio quasars, simulations need to produce disks with high radiative efficiency, as well as high jet efficiency. The quasar-like radiation from the nucleus appears to require a scaleheight of 0.01 from analytic models. Achieving this has been a challenge. The decrease in disk efficiency associated with an increase in jet efficiency was thought to be a good description of hard state jets in X-ray binaries and of radio galaxies, both observed to produce X-ray and radio jet signatures but weak emission lines. But radio quasars appear to upend this inverse trend between jet and disk efficiency. Ref.~\cite{30} have explored the blazar subclass of radio quasars (flat spectrum radio quasars (FSRQs), attempting to fit observations with simulations of disks in the context of moderately thin disks (those of~\cite{25}). We will explore this in the next section but the basic conclusion is the need for some missing ingredient that allows for both large jet and disk efficiency. Finally, Ref.~\cite{16} have produced the thinnest disks in GRMHD, with $H/R \sim 0.03$, with the absence of jet suppression. This is an important milestone for simulations. Ref.~\cite{16,17} also attempt to explain radio quasars, which they explore in the context of scale invariance, anchoring their arguments to transitory ballistic jets observed in X-ray binaries. We explore their analysis in that context, as well.

\subsection{FSRQ Jets from Moderately Thin Disks}
Flat-spectrum radio quasars (FSRQs) are the blazar subclass of active galactic nuclei (AGN) emitting the most relativistic jets along our line of sight with relatively high accretion rates. Using gamma-ray luminosity from the Fermi Large Area Telescope as a proxy for jet powers, and independent measurements of black hole mass, Ref.~\cite{30} produce the best fit correlation between jet power and black hole mass from observations of 154 FSRQs. The ideas are built on the work of~\cite{31} who uncovered a
correlation between gamma-ray luminosity $L_{\gamma}$ and total blazar jet power $P_{jet}$ given by
\begin{equation}
log_{10} P_{jet} = 0.51 log_{10}L_{\gamma} + 21.2,
\end{equation}
in units of erg/s. Soares \& Nemmen (2020) compiled observational data for 154 FSRQs and fitted a relation between $L_{\gamma}$ and black hole mass M, obtaining
\begin{equation}
log_{10} M = 0.37 log_{10}L_{\gamma} - 8.95,
\end{equation}
with $M$ in terms of solar masses and $L_{\gamma}$ in erg/s. Combining Equations (1) and (2), we~obtain
\begin{equation}
log_{10} P_{jet} = 1.38 log_{10}M + 33.53,
\end{equation}
which produces constraints on models that connect jet power to black hole mass. Ref.~\cite{30} show that $97\%$ of their FSRQ sample satisfies jet energetics as a function of black hole mass (Equation (3)) for moderately thin disks ($H/R=0.13$). If the radiative efficiency in such moderately thin disks is insufficient to explain the quasar-like spectrum, the success of equation (3) to the sample becomes moot. In that case, it means the
scaleheight must satisfy $H/R < 0.1$, which means the GRMHD simulations adopted by~\cite{30} fail to explain the sample. If, as mentioned, radiatively efficient disks satisfy $H/R$ = 0.01, the mismatch is of course much worse. Ref.~\cite{30} report that it drops from $0.098$ to $6 \times 10^{-4}$, which is 163 times smaller. The basic question here is whether or not the disk thickness that is compatible with jet energetics, is also compatible with the required radiative efficiency. What has been a constant staple in GRMHD simulations is the trend noted above. The thickness that seems to be needed to explain the high enough jet efficiency works against the thinness that is needed to explain the radiative output. A common way out of the above conundrum has been to assume that radio quasars do not have radiatively efficient disks into the inner regions and that while further out the radiation escapes, the energy is retained in the inner regions and the disk scaleheight puffs up there as a result. And this thick disk would be needed to explain the formation of a jet. This, it was thought for decades, would explain a quasar-like spectrum due to the outer cool, radiatively efficient disk, coupled to a powerful jet produced by a thick inner disk. Soares \& Nemmen (2019) also mention this possibility. Several recent observations show that this idea fails (\cite{32,33,34,35} and references therein).

While magnetically arrested GRMHD disk simulations struggle from a numerical perspective to accurately evolve strongly magnetized regions, radiation is arguably a more difficult problem. As a result, it is treated with a variety of prescriptions from ad hoc cooling functions to keep the disk as thin as possible (e.g.,~\cite{25}) to more realistic ones that include radiative transport~\cite{36}. Given the differences in how the disk is cooled, the radiative nature of the disk can vary greatly among simulations, giving little confidence that GRMHD is providing insights into radiatively efficient disk physics. In~\cite{25}, for example, the efficiency is twice that of Novikov \& Thorne, but it is all emitted within the very central region. This appears difficult to square with typical quasar spectra.

A promising result is obtained in the GRMHD simulations of [16], with the GPU
-accelerated code H-AMR 
\cite{37}, which simulates the thinnest possible tilted disk with good enough resolution to resolve the magnetorotational instability with $H/R =0.03$ around a black hole of spin $0.9375$. This simulation also provided a demonstration within a turbulent MHD accretion disk of the Bardeen-Petterson alignment of the disk and black hole. The total efficiency of the disk was found to be higher than that of the Novikov \& Thorne prediction by a factor of 3. Perhaps most importantly, a powerful jet was present.

\subsection{Jet and Disk Efficiency in GRMHD}
\textls[-15]{Attempts at understanding the compatibility and/or difference between GRMHD and their ability to model thin disks and their compatibility with jets to model radio quasars are obscured in that some GRMHD simulations incorporate or produce physics that is completely different from others. For example, some simulations argue that the Novikov \& Thorne solution is not too different from what the disk produces in GRMHD (e.g.,~\cite{29,38} while other more recent simulations with magnetic fields flooding the system (e.g.,~\cite{25,39,40,41}} conclude that almost all the radiation is emitted near or inside of the innermost stable circular orbit (ISCO). Other GRMHD simulations suggest that strong magnetic fields might be generated in situ via dynamo action~\cite{42}. What is the origin of such vastly different physical pictures? One clear candidate is the variety of different mechanisms incorporated in simulations to treat radiation, with some ad-hoc while others are more faithful to the physics. For example, Ref.~\cite{36} performed 3D GRMHD simulations taking into account the time-dependent radiative transfer equations of accretion disks with $H/R \sim 0.1$ around a black hole with a spin of 0.5 using the HARMRAD code~\cite{43}. The radiative efficiency of the disk was found to be slightly lower than that of the Novikov \& Thorne prediction while in the case of~\cite{25}, the efficiency was twice the efficiency of Novikov \& Thorne when the effects of scattering and absorption of radiation were not taken into account. These different processes change the efficiency but also the location where the radiation is emitted, as already discussed. Ref.~\cite{44} performed a radiative GRMHD simulation of a geometrically thin, sub-Eddington accreting disk ($H/R \sim 0.15$) around a zero spin black hole using the KORAL code~\cite{45,46}. Although a significant amount of dissipation was found inside the marginally stable orbit as in~\cite{25} simulations, the efficiency of the disk was found to be very close to that of Novikov \& Thorne. {All the results from available GRMHD simulations have been summarized in Table \ref{tab1}.}

Uncertainty very much still dominates our understanding of numerical simulations of accretion onto black holes, but a bottom line is beginning to emerge: Strong disk magnetic fields seem to be a by-product of strong black hole threading magnetic fields but strong disk magnetic fields do not allow disks to remain thin~\cite{47,48,49}. And disks that are not thin might not be able to radiate efficiently. It seems there is possibly another factor that can allow the compatibility between the thinness of the disk and the condition of jet formation.

\subsection{Scale Invariance and the Jet-Disk Connection}
Let us assume that jet efficiency is sufficient in GRMHD to explain radio quasars. We will now show that arguments explaining radio quasars using these simulations nonetheless run into problems. As discussed by~\cite{16}, radiatively efficient disks appear not to produce relativistic jets, which suggests to them that they are uncovering a transitional process that manifests itself in black hole X-ray binaries as the disk evolves from the hard state to the soft state~\cite{7}. The first issue here is the need to show in {GRMHD} that as the disk collapses from the hard state and transitions toward the soft state, the jet power depends on black hole spin, whereas in the state that precedes it (the hard state), the jet power is less dependent on black hole spin. In other words, it is crucial that this time-dependent scenario must be simulated. These conclusions are based on the observation that jet power does not correlate with a spin in the hard state~\cite{9}, while it may correlate with it in the transitory state~\cite{13,50}. Second, and more fundamentally, the time evolution implicit in the above scenario does not lend itself to scaling up this transitory jet and applying it to radioquasars~\cite{51,52}. This crucial point has by and large gone unnoticed. The problem is that radio quasars distribute themselves on average at higher redshift compared to FRI radio galaxies and such a distribution cannot emerge from transitions that take hard state jets into transitory ballistic jets, which reverse the time sequence. The scale-invariant approach, thus, predicts radio quasars occupying lower redshifts compared to FRI radio galaxies. If we incorporate the full cyclical behavior of X-ray binaries in a scale-invariant application to radio quasars, then we would conclude that radio quasars and FRI radio galaxies do not show any redshift dependent difference, but that best-case scenario is not observed. In short, using the time evolution of X-ray binaries in trying to understand radio quasars predicts either that radio quasars are distributed at relatively lower redshifts or that they are not distributed differently in redshift compared to FRI radio galaxies. But the observations are not compatible with these scenarios.

%

\begin{specialtable}[H]
\caption{Results from some representative GRMHD simulation works showing black hole spin (a/M), scale height (H/R), disk radiative efficiency ($\eta_{r}$), jet efficiency ($\eta_{j}$ ), total efficiency ($\eta_{T}$) and cooling method implemented. Here, $\eta_{NT}$ means efficiency predicted by the Novikov Thorne (1973) model and NA means not available information.\label{tab}}
\begin{tabular}{|p{1.9cm}|p{1.5cm}|p{1.5cm}|p{1.5cm}|p{1.5cm}|p{1.5cm}|p{1.5cm}|}

\toprule
\textbf{References}    & \textbf{a/M}    & \textbf{H/R}    & \textbf{$\eta_{r}$}    & \textbf{$\eta_{j}$}    & \textbf{$\eta_{T}$}    & \textbf{Cooling}\\
\midrule
Penna et al. (2010)    & 0-0.98    & 0.07-0.3    & Less than 4.5 $\%$ deviation from $\eta_{NT}$    & NA    & Close to $\eta_{NT}$    & Ad hoc as in Shafee et al. (2008)\\
\midrule
McKinney, Tchekhovskoy \& Blandford (2012)    & -0.9375 - 0.99    & 0.2 - 1    & NA    & Up to 50 times higher than $\eta_{NT}$    & Up to 120 times higher than $\eta_{NT}$    & No\\
\midrule
Avara, McKinney \& Reynolds (2016)    & 0.5    & 0.05-1    & $15\%$ (almost 2 times higher than $\eta_{NT}$ = $8.2\%$    & $1\%$ (almost an order less than $\eta_{NT}$)    & $20\%$ (around 2.5 times higher than $\eta_{NT}$)    & Ad hoc as in Noble et al. (2010)\\
\midrule
Sadowski (2016)    & 0    & 0.15    & 5.5$\pm$0.5$\%$ (very close to $\eta_{NT}$ = 5.7 $\%$)    & NA    & NA    & Radiative transfer\\
\midrule
Morales Texeira, Avara \& McKinney (2018)    & 0.5    &    0.1    & 2.9$\%$ (less than half of $\eta_{NT}$ = 8.2 $\%$)    & 4.3$\%$ (around half of $\eta_{NT}$)    & 18.6$\%$(more than twice of $\eta_{NT}$)    & Radiative transfer\\
\midrule
Liska et al. (2019)    & 0.9375    &    0.03 & 18$\%$ (close to $\eta_{NT}$ = 17.9$\%$)    & 20-50$\%$ (Up to 2.5 times higher than $\eta_{NT}$)    & 60-80$\%$(3-4 times higher than $\eta_{NT}$)    & Ad hoc as in Noble et al. (2010)\\

\bottomrule
\end{tabular}\label{tab1}
\end{specialtable}

These issues---and many others---have been resolved in semi-analytic models~\cite{4} via the introduction of a key feature for accreting black holes, namely counterrotation between the disk and the black hole, a window on accretion than many have found fruitful (a subset of these are~\cite{53,54,55,56,57,58,59,60,61,62,63,64,65,66,67,68}. In counter-rotating disk configurations, the process that suppresses the jet weakens as the black hole spin increases, thereby allowing the conditions that lead to strong, collimated jets, to couple to bright disk states. The radio-loud/radio-quiet dichotomy in this picture is based on a high black hole spin for both families of accreting black holes. Whereas radio quasars are high spinning counter-rotating accretion configurations, jetless quasars are high spinning, prograde accreting black holes. This is due to the small value of the inner disk boundary for high prograde configurations. Because the disk reaches deep into the gravitational potential of the black hole for high prograde spins, a greater amount of energy is generated from the disk at all disk locations, and the disk is radiatively dominated or quasar-like. Recently,  we have been able to explain the lack of symmetry in the radio loud/radio quiet dichotomy with jetless (or radio quiet) AGN dominating the distribution at about 80\%~\cite{33}.
These ideas have allowed for theory to be compatible with scale invariance and with the observed distribution of radio-loud and radio-quiet quasars across cosmic time. Our roadmap for GRMHD simulations involves, therefore, exploring the nature of the jet-disk connection for corotating, as well as counterrotating accreting black holes. Whereas the jet efficiency was found to be slightly higher in the former, we suspect that, with better treatment of the radiation processes, the results will start to shift, and counterrotation will emerge as a key ingredient.

\section{Conclusions}
The existence of a subset of active galaxies with high observed radiative efficiency and powerful jets---radio quasars---has remained a major unsolved problem in high energy astrophysics for decades. In this work, we have explored the state of the art in numerical simulations of accretion onto black holes and find that where the comparison to observation is possible, tensions arise. We find a variety of very different physical scenarios that fail to produce a coherent picture for radio quasars and their FSRQ subclass from GRMHD. And we have singled out what appears to be the problem: the strong disk magnetic fields that accompany the strong black hole-threading field work to support greater disk thickness, which in turn decreases the radiative efficiency of the disk. How can we simulate accreting black holes capable of sustaining strong black hole horizon magnetic fields despite weak magnetic fields in their disk? The simulations of~\cite{16,17} constitute a milestone in this respect since they produce strong jets despite unprecedented small $H/R$ values. Much work is required to single out the physical processes that would allow such disks to still drag the sufficiently strong magnetic field onto the black hole despite the disk thinness. And, this would need to be understood and juxtaposed to simulations where increased disk thinness has the opposite effect on the magnetic field threading the black hole, or on some other quantity associated with jet suppression. Whatever scale-invariant processes will be identified, it is clear that they cannot be associated with transitory ballistic jets in X-ray binaries. We have then made contact with the idea that counterrotating black holes may resolve such issues and have thus encouraged the GRMHD community to go back and explore the jet-disk connection in that context with these new simulations.



\vspace{6pt}



\authorcontributions{
Conceptualization, D.G. and C.B.S.; methodology, D.G.; validation, D.G., C.B.S., and B.L.; investigation, D.G., C.B.S., and B.L.; writing---original draft preparation, D.G. and C.B.S.; writing---review and editing, D.G. and C.B.S.; supervision, D.G.; project administration, D.G.; funding acquisition, D.G.and C.B.S. All authors have read and agreed to the published version of the~manuscript.}

\funding{This research was funded by the National Natural Science Foundation of China grant number 12073021.}



\dataavailability{{No data was generated during this work.}}


\conflictsofinterest{The authors declare no conflict of interest. }
\end{paracol}
\reftitle{References}





\end{document}